\documentclass[runningheads]{llncs}

\usepackage[T1]{fontenc}
\usepackage{graphicx}
\usepackage{amsfonts}
\usepackage{bbm}
\usepackage[hidelinks]{hyperref}
\usepackage{float}
\usepackage{paralist}
\usepackage{array}
\usepackage{adjustbox}
\newcolumntype{H}{>{\setbox0=\hbox\bgroup}c<{\egroup}@{}}
\usepackage{xcolor}

\newcommand{\qqueries}{\hat q^{queries}}
\newcommand{\qanswers}[1][n,k]{\hat q^{answers}_{#1}}

\newcommand{\qrep}{\hat q}
\newcommand{\drep}{\hat d}
\newcommand{\CLS}{\mathtt{[CLS]}}
\newcommand{\SEP}{\mathtt{[SEP]}}

\begin{document}

\title{CoSPLADE: Contextualizing SPLADE   for Conversational Information Retrieval}
\titlerunning{CoSPLADE: Contextualizing SPLADE   for Conversational IR}

\author{
Nam Le Hai\inst{1}\orcidID{0000-0002-9020-8790}
\and Thomas Gerald\inst{2}
\and Thibault Formal\inst{1,3}
\and Jian-Yun Nie\inst{4}
\and Benjamin Piwowarski\inst{1}\orcidID{0000-0001-6792-3262}
\and Laure Soulier\inst{1,2}\orcidID{0000-0001-9827-7400}
}

\institute{
Sorbonne Université, CNRS, ISIR, F-75005 Paris, France \email{\emph{first.last}@sorbonne-universite.fr} \and
Université Paris-Saclay, CNRS, SATT Paris Saclay, LISN, 91405 Orsay, France \email{\emph{first.last}@lisn.upsaclay.fr} \and
Naver Labs Europe, Meylan, France \email{\emph{first.last}@naverlabs.com} \and
University of Montreal, Montreal, Canada \email{nie@iro.umontreal.ca}
}

\authorrunning{Le Hai et al.}

%


\maketitle              

\begin{abstract}
Conversational search is a difficult task as it aims at retrieving documents based not only on the current user query but also on the full conversation history.
Most of the previous methods have focused on a multi-stage ranking approach relying on query reformulation, a critical intermediate step that might lead to a sub-optimal retrieval. Other approaches have tried to use a fully neural IR first-stage, but are either zero-shot or rely on full learning-to-rank based on a dataset with pseudo-labels. In this work, leveraging the CANARD dataset, we propose an innovative lightweight learning technique to train a first-stage ranker based on SPLADE. By relying on SPLADE sparse representations, we show that, when combined with a second-stage ranker based on T5Mono, the results are competitive on the TREC CAsT 2020 and 2021 tracks. The source code is available at https://github.com/xpmir/cosplade

\keywords{information retrieval   \and conversational search \and first-stage ranking.}

\end{abstract}

\section{Introduction}

With the introduction of conversational assistants like Siri, Alexa or Cortana, conversational Information Retrieval, a variant of adhoc IR, has emerged as an important research domain \cite{Culpepper0S18,dalton_trec_2019}. In conversational IR, a search is conducted within a session, and the user's information need is expressed through a sequence of queries, similarly to natural conversations -- thus introducing complex inter-dependencies between queries and responses.

Not surprisingly, neural IR models have been shown to perform the best on conversational IR~\cite{dalton_cast_2020,dalton_cast_2021}. Most prior works rely on a Historical Query Expansion step~\cite{ZamaniConversationalInformationSeeking2022}, i.e. a query expansion mechanism that takes into account all past queries and their associated answers. Such query expansion model is learned on the CANARD dataset \cite{ElgoharyCanYouUnpack2019}, which is composed of a series of questions and their associated answers, together with a disambiguated query, referred to as \emph{gold query} in this paper. However, relying on a reformulation step is computationally costly and might be sub-optimal as underlined in~\cite{KrasakisZeroshotQueryContextualization2022b,lin_contextualized_2021}. Krasakis et al. \cite{KrasakisZeroshotQueryContextualization2022b} proposed to use ColBERT~\cite{khattab_colbert_2020} in a zero-shot manner, replacing the query by the sequence of queries, without any training of the model. Lin et al. \cite{lin_contextualized_2021} proposed to learn a dense \emph{contextualized} representation of the query history, optimizing a learning-to-rank loss over a dataset composed of weak labels. This makes the training process complex (labels are not reliable) and long.

In this work, we follow this direction of research but propose a much lighter training process for the first-stage ranker, where we focus on queries and do not make use of any passage -- and thus of a learning-to-rank training. It moreover sidesteps the problem of having to derive weak labels from the CANARD dataset\footnote{Note that for the second stage, we rely on weak labels since our model is similar to previous works. Given that the gap between first-stage and second-stage rankers continues to decrease, training a second-stage ranker might not be necessary in the future.}.
Given this strong supervision, we can consider more context -- i.e. we use the answers provided by the system the user is interacting with, which allows to better contextualize the query, as shown in our experiments.
The training loss we propose leverages the sparse representation of queries and documents provided by the SPLADE model~\cite{formal_splade_v2}.
In a nutshell, we require that the representation of the query matches that of the disambiguated query (i.e. the \emph{gold query}). 
Our first-stage ranker achieves high performances, especially on recall -- the most important measure in a multi-stage approach, comparable to the best systems in TREC CAsT~\cite{dalton_cast_2021}, but also on precision-oriented measures -- which shows the potential of our methodology.




Finally, to perform well, the second-stage ranker (i.e. re-ranker) needs to consider the conversation as well, which might require a set of heuristics to select some content and/or query reformulation such as those used in \cite{LinTREC2020Notebook2021}. Leveraging the fact that our first-stage ranker outputs weights over the BERT vocabulary, we propose a simple mechanism that provides a conversational context to the re-ranker in the form of keywords selected by SPLADE.

%
In summary, our contributions are the following:
\begin{enumerate}
\vspace{-0.2cm}
    \item We propose the CoSPLADE (COntextualized SPLADE) model based on a new loss to optimize a first-stage ranker resulting in a lightweight training strategy and state-of-the-art results in terms of recall;
    \item We show that, when combined with a second-stage ranker based on a context derived from the SPLADE query representation of the first stage, we obtain results on par with the best approaches in TREC CAsT 2020 and 2021.
\end{enumerate}




\vspace{-0.4cm}
\section{Related Works}

The first edition~\cite{dalton_cast_2020} of the TREC Conversational Assistance Track (\textit{CAsT}) was implemented in 2019, providing a new  challenge on Conversational Search. The principle is the following: a user queries the system with questions in natural language, and each time gets a response from the system. The challenge differs from classical search systems as involving previous utterances (either queries or answers) is key to better comprehending the user intent.
In conversational IR, and in TREC \textit{CAsT} \cite{dalton_trec_2019,dalton_cast_2020,dalton_cast_2021} in particular, the sheer size of the document collection implies to design an efficient (and effective) search system. 

Conversational IR is closely related to conversational Question-Answering \cite{bert2019chen,reddy-etal-2019-coqa,qu2019attentive} in the sense that they both include interaction turns in natural language. However, the objective is intrinsically different. While the topic or the context (i.e., the passage containing answers) is known in conversational QA, conversational IR aims to search among a huge collection of documents with potentially more exploratory topics. 
With this in mind, in the following, we focus on the literature review of conversational IR.

We can distinguish two lines of work in conversational search.
The first one \cite{voskarides_ilps_2019,voskarides_query_2020,Yang2019QueryAA,waterloo2019clarke} focuses  on a Contextual Query Reformulation (CQR) to produce a (plain or bag-of-words) query, representing ideally the information need free of context, which is fed into a search model. One strategy of CQR consists in selecting keywords from previous utterances by relying on a graph weighted by either word2vec similarity \cite{voskarides_ilps_2019}, term-based importance using BM25 \cite{lin_multi-stage_2021}, or classification models \cite{voskarides_query_2020}.
Other approaches \cite{kumar_making_nodate,lin_multi-stage_2021,LinTREC2020Notebook2021,yu_few-shot_2020,vakulenko_question_2021} leverage the potential of generative language models (e.g., GPT2 or T5) to rewrite the query. Such approaches are particularly effective, reaching top performances in the TREC CAsT 2020 edition \cite{dalton_cast_2020}.
Query reformulation models also differ in the selected evidence sources. Models either focus on the early stage of the conversation \cite{aliannejadi_harnessing_2020}, on a set of the queries filtered either heuristically \cite{waterloo2020} or by a classification model \cite{mele_finding_2021}, or on both previous queries and documents~\cite{waterloo2021}. 
Finally, to avoid the problem of generating a \emph{single} query, \cite{kumar_making_nodate,query2020lin} have proposed to use different generated queries and aggregate the returned documents.

The reformulation step is however a bottleneck since there is no guarantee that the ``gold query'' is optimal and thus generalizes well~\cite{lin_contextualized_2021,KrasakisZeroshotQueryContextualization2022b}. Moreover, generating text is time-consuming.
To avoid these problems, the second line of work aims to directly integrate the conversation history into the retrieval model, bypassing the query reformulation step.  
As far as we know, only a few studies followed this path in conversational search.
Qu et al~\cite{open2020qu} compute a query representation using the $k$ last queries in the dialogue~\cite{albert2020Lan}.
Similarly, Lin et al.~\cite{lin_contextualized_2021} average contextualized tokens embeddings over the whole query history. The representation is learned by optimizing a learning-to-rank loss over a collection with weak labels, which requires much care to ensure good generalization.
Finally, Krasakis et al.~\cite{KrasakisZeroshotQueryContextualization2022b} use a more lexical neural model, i.e. ColBERT~\cite{khattab_colbert_2020}, to encode the query with its context -- but they do not finetune it at all.
In this work, we go further by using a sparse model SPLADE~\cite{formal_splade_v2}, using a novel loss tailored to such sparse representations, and by using a lightweight training procedure that does not rely on passages, but only on a dataset containing reformulated queries.

\vspace{-0.2cm}
\section{Model}


In TREC CAsT \cite{dalton_cast_2020,dalton_cast_2021}, each retrieval session contains around 10 turns of exchange. 
Each turn corresponds to a query and its associated canonical answer\footnote{Selected by the organizer as the most relevant answer of a baseline system.} is provided as context for future queries. 
Let us now introduce some notations that we use to describe our model.
For each turn $n \le N$, where $N$ is the last turn of the conversation, we denote by $q_n$ and $a_n$ respectively the corresponding query and its response. Finally, the context of a query $q_n$ at turn $n$ corresponds to all the previous queries and answers, i.e.
 $q_1,\ a_1,\ q_2,\ a_2,\ ...,\ q_{n-1},\ a_{n-1}$.
The main objective of the \textit{TREC CAsT} challenges is to retrieve, for each query $q_n$ and its context, the relevant passages. 
In the next sections, we present our first-stage ranker and second-stage re-ranker, along with their training procedure,
both based, directly or indirectly, on the SPLADE (v2) model described in~\cite{formal_splade_v2}. SPLADE has shown results on par with dense approaches on in-domain collections while exhibiting stronger abilities to generalize in a zero-shot setting~\cite{formal_splade_v2}. Moreover, it outputs a sparse representation of a document or a query in the BERT vocabulary, which is key to our model during training and inference. This explains why we did not consider interaction models such as ColBERT \cite{santhanam-etal-2022-colbertv2} or dense approaches \cite{10.1145/3404835.3462891}.
The SPLADE model we use includes a contextual encoding function, followed by some aggregation steps: ReLU, log saturation, and max pooling over each token in the text. The output of SPLADE is a sparse vector with only positive or zero components in the BERT vocabulary space $\mathbb{R}^{|V|}$. In this work, we use several sets of parameters for the same SPLADE architecture and distinguish each version by its parameters $\theta$, and the corresponding model by $SPLADE(\ldots;\theta)$.

\subsection{First stage}
\label{sec:first-stage}

The original SPLADE model \cite{formal_splade_v2} scores a document using the dot product between the sparse representation of a document ($\drep$) and of a query ($\qrep$):
 \begin{equation}
 \label{eq:rsv}
 s(\qrep, \drep) = \qrep \cdot \drep
 \end{equation}
 In our work, like in \cite{lin_contextualized_2021}, we suppose that the document representation has been sufficiently well-tuned on the standard ad-hoc IR task. The document embedding $\drep$ is thus obtained using the pre-trained SPLADE model, i.e. $\drep = SPLADE(\CLS\ d;\ \theta_{SPLADE})$ where $\theta_{SPLADE}$ are the original SPLADE parameters obtained from HuggingFace\footnote{The weights can be found at \url{https://huggingface.co/naver/splade-cocondenser-ensembledistil}}. These parameters are not fine-tuned during the training process.  We can thus use standard index built from the original SPLADE document representations to retrieve efficiently the top-$k$ documents. In the following, we present how to contextualize the query representation using the conversation history. Then,  we detail the training loss aiming at reducing the  gap between the  representation of the gold query and the contextualized representation.

\paragraph{Query representation.}

Like state-of-the-art approaches for first-stage conversational ranking \cite{lin_contextualized_2021,KrasakisZeroshotQueryContextualization2022b}, we contextualize the query with the previous ones. Going further, we propose to include the answers in the query representation process, which is easier to do thanks to our lightweight training.

To leverage both contexts, we use a simple model where the contextual query representation at turn $n$, denoted by $\qrep_{n,k}$, is the combination of two representations, $\qqueries_n$ which encodes the current query in the context of all the previous queries, and $\qanswers$ which encodes the current query in the context of $k$ the past answers\footnote{In the experiments, we also explore an alternative model where answers and queries are considered at once.}. 
Formally, the contextualized query representation $\qrep_{n,k}$ is:
 \begin{equation}
 \label{eq:qrep-full}
 \qrep_{n,k} = \qqueries_n + \qanswers
 \end{equation}
where we use two versions of SPLADE parameterized by $\theta_{queries}$ for the full query history and $\theta_{answers,k}$ for the answers. These parameters are learned by optimizing the loss defined in Eq. (\ref{eq:full-loss}).

Following \cite{lin_contextualized_2021}, we define $\qqueries_n$ to be the query representation produced by encoding the concatenation of the current query and all the previous ones:
 \begin{equation}
 \label{eq:qrep-hist}
\qqueries_n = SPLADE(\CLS\ q_n\ \SEP\ q_1\ \SEP\ \ldots\ \SEP\  q_{n-1}; \theta_{queries})
 \end{equation}
using a set of specific parameters $\theta_{queries}$.

To take into account the answers that the user had access to, we need to include them in the representation. Following prior work \cite{waterloo2020}, we can consider a various number of answers $k$, and in particular, we can either choose $k=1$ (the last answer) or $k=n-1$ (all the previous answers). Formally, the representation $\qanswers$ is computed as the mean of the representations of query-answer pairs. This allows to evade the token-length limits imposed by language models. Formally,
\begin{equation}
\label{eq:qanswers}
\qanswers = \frac{1}{k}  \sum_{i = n-k}^{n-1} SPLADE(q_n\ \SEP\ a_i; \theta_{answers,k})
\end{equation}

\paragraph{Training}

Based on the above, training aims at obtaining a good representation $\qrep_n$ for the last issued query $q_n$, i.e. to contextualize $q_n$ using the previous queries and answers. To do so, we can leverage the gold query $q^*_n$, that is (hopefully), a  contextualized and unambiguous query. We can compute the representation $\hat q^*_n$ of this query by using the original SPLADE model, i.e.
 \begin{equation}
 \label{eq:gold-q-splade}
 \qrep_n^* = SPLADE(q_n^*; \theta_{SPLADE})
 \end{equation}
  
For example, for a query "How old is he?" the matching gold query could be "How old is Obama?". The representation of the latter given by SPLADE would be as follows:
 \[
 [("Obama",1.5), ("Barack",1.2), ("age",1.2), ("old",1.0), ("president",0.8),...] \]
where the terms ``Obama''  and ``Barack'' clearly appear alongside other words related to the current query (``old'' and the semantically related ``age'').

We can now define the goal of the training, which is to reduce the difference between the gold query representation $\qrep_n^*$ and the representation $\qrep_{n,k}$ computed by our model.
An obvious choice of a loss function is to match the predicted and gold representations using cosine loss (since the ranking is invariant when scaling the query). However, as shown in the result section, we experimentally found better results with a modified MSE loss, whose first component is the standard MSE loss:
 \begin{equation}
 \label{eq:mse-loss}
 Loss_{MSE}(\qrep_{n,k}, \qrep_n^*) = MSE(\qrep_{n,k}, \qrep_n^*)
 \end{equation}
 
In our experiments, we observed that models trained with the direct MSE do not capture well words from the context, especially for words from the answers. 
The reason is that the manually reformulated gold query usually only contains a few additional words from the previous turns that are directly implied by the last query. Other potentially useful words from the answers may not be included. This is a conservative expansion strategy which may not be the best example to follow by an automatic query rewriting process.
We thus added an asymmetric MSE, designed to encourage term expansion from past answers, but avoid introducing noise by restricting the terms to those present in the gold query $q^*_n$. Formally, our asymmetric loss is:
 \begin{equation}
  \label{eq:asymloss}
 Loss_{asym}(\qanswers, \qrep_n^*) = 
 \left(
    \max(
        \qrep_n^* - \qanswers,
        0
    ) 
\right)^2
 \end{equation}
 where the maximum is component-wise. This loss thus pushes the answer-biased representation $\qanswers$ to include tokens  from the gold query. Contrarily to MSE, it does not impose (directly) an upper bound on the components of the $\qanswers$ representation -- this is done indirectly through the final loss function described below.

The final loss we optimize is a simple linear combination of the losses defined above, and only relies on computing two query representations:
 \begin{equation}
 \label{eq:full-loss}
 Loss(\qrep_{n,k}, \qrep_n^*) = Loss_{MSE}(\qrep_{n,k}, \qrep_n^*) +  Loss_{asym}(\qanswers, \qrep_n^*)
 \end{equation}
There is an interplay between the two components of the global loss. More precisely, $Loss_{asym}$ pushes the $\qanswers$ representation to match the golden query representation $\qrep_n^*$ \emph{if it can}, and $Loss_{MSE}$ pushes the queries-biased representation $\qrep_{n,k}$ to compensate \emph{if not}. It thus puts a strong focus on extracting information from past answers, which is shown to be beneficial in our experiments. 

\paragraph{Implementation details.}
For the first-stage, we initialize both encoders (one encoding the queries, and the other encoding the previous answer) with pre-trained weights from SPLADE model for adhoc retrieval. We use the ADAM optimizer with train batch size 16, learning rate 2e-5 for the first encoder and 3e-5 for the second. We fine-tune for only 1 epoch over the CANARD dataset. 

\subsection{Reranking}

We perform reranking using a T5Mono~\cite{nogueira_document_2020} approach,
where we enrich the raw query $q_n$ with keywords identified by the first-stage ranker. Our motivation is that these words should capture the information needed to contextualize the raw query. The enriched query $q_n^+$ for conversational turn $n$ is as follows:
\begin{equation}
\label{eq:reranking:t5-contextual-prompt}
q_n^+ = q_n.\ Context: q_1\ q_2\ \ldots\ q_{n-1}.\ Keywords: w_1, w_2, ..., w_{K}
\end{equation}
where the $w_i$ are the top-$K$ most important words that we select by leveraging the first-stage ranker as follows. First, to reduce noise, we only consider words that appear either in any query $q_i$ or in the associated answers $a_i$ (for $i\le n-1$). 
Second, we order words by using the maximum SPLADE weight over tokens that compose the word.\footnote{To improve coherence, we chose to make keywords follow their order of appearance in the context, but did not vary this experimental setting.}

We denote the T5 model fine-tuned for this input as $T5^+$. 
As in the original paper~\cite{nogueira_document_2020}, the relevance score of a document $d$ for the query $q_n$ is the probability of generating the token ``\texttt{true}'' 
given a prompt pt$(q_n^+,d)=$ 
``\texttt{Query: }$q_n^+$. \texttt{Document: }$d$\texttt{. Relevant:}'':
\begin{equation}
\label{eq:reranking-t5-score}
score(q_n^+,d; \theta) = \frac
    {
        p_{T5}( \mathtt{true} |\textrm{pt}(q_n^+, d); \theta)
    }
    {
        p_{T5}(\mathtt{true}| \textrm{pt}(q_n^+, d); \theta) 
        + p_{T5}(\mathtt{false} | \textrm{pt}(q_n^+, d); \theta)
    }
\end{equation}
where $\theta$ are the parameters of the T5Mono model.

Differently to the first stage training, we fine-tune the ranker by aligning the scores of the documents, and not the weight of a query (which is obviously not possible with the T5 model). Here the ``gold'' score of a document is computed using the original T5Mono with the gold query $q^*_n$. The T5 model is initialized with weights made public by the original authors\footnote{We used the Huggingface checkpoint \url{https://huggingface.co/castorini/monot5-base-msmarco}}, denoted as $\theta_{T5}$. 
More precisely, we finetune the pre-trained T5Mono model using the MSE-Margin loss~\cite{Hofsttter2020ImprovingEN}. The loss function for the re-ranker (at conversation turn $n$, given documents $d_1$ and $d_2$) is computed as follows:
 \begin{eqnarray*}
 \mathcal L_R &=& \left[ \left(s(q_n^+,d_1; \theta_{T5+}) - s(q_n^+,d_2; \theta_{T5+})\right) - \left( s(q_n^*,d_1; \theta_{T5}) - s(q_n^*,d_2; \theta_{T5})  \right) \right]^2
 \end{eqnarray*}
We optimize the $\theta_{T5+}$ parameters by keeping the original $\theta_{T5}$ to evaluate the score of gold queries.

\paragraph{Implementation details.}
We initialize $\theta_{T5+}$ as $\theta_{T5}$, and fine-tune for 3 epochs, with a batch size of 8 and a learning rate 1e-4. We sample pairs $(d_1,d_2)$ using the first-stage top-1000 documents: $d_1$ is sampled among the top-3, and $d_2$ among the remaining 997 to push the model to focus on important differences in scores.


\section{Experimental Protocol}
We designed the evaluation protocol to satisfy two main evaluation objectives:
\begin{inparaenum}[(i)]
    \item Evaluating separately the effectiveness of the first-stage and the second-stage ranking components of our CoSPLADE model;
    \item Comparing the effectiveness of our CoSPLADE model with TREC CAsT 2020 and 2021 participants.
\end{inparaenum}

\subsection{Datasets}

To train our model, we used the CANARD corpus\footnote{\url{https://sites.google.com/view/qanta/projects/canard}}, a conversational dataset focusing on context-based query rewriting. More specifically, the CANARD dataset is a list of conversation histories, each being composed of a series of queries, short answers (human-written), and reformulated queries (contextualized).  The training, development, and test sets include respectively $31.538$, $3.418$, and $5.571$ contextual and reformulated queries.

To evaluate our model, we used the TREC CAsT 2020 and 2021 datasets which include respectively 25 and 26 information needs (topics) and a document collection composed of the  MS MARCO dataset, an updated dump of Wikipedia from the KILT benchmark, and the Washington Post V4 collection.
For each topic, a conversation is available, alternating questions and responses (manually selected passages from the collection, aka canonical answers). For each question (216 and 239 in total), the dataset provides its manually rewritten form as well as a set of about 20 relevant documents.  We use the former to define an upper-bound baseline (\textbf{Splade\_GoldQuery}). 

\subsection{Metrics and baselines}
We used the official evaluation metrics considered in  the TREC CAsT 2020 and 2021,  namely nDCG@3, MRR, Recall@X, MAP@X, nDCG@X, where the cut-off is set to 1000 for the CAsT 2020 and 500 for the CAsT 2021. For each metric, we calculate the mean and variance of performance across the different queries in the dataset.
With this in mind, we present below the different baselines and scenarios used to evaluate each component of our model.

\subsubsection{First-stage ranking scenarios.}

To evaluate the effectiveness of our first-stage ranking model (Section \ref{sec:first-stage}), we  compare our approach CoSPLADE, based on the query representation of Eq. (\ref{eq:qrep-full}) with different variants (the document encoder is set to the original SPLADE encoder throughout our experiments):
\begin{inparadesc}
\item[SPLADE\_rawQuery] (lower bound):  SPLADE \cite{formal_splade_2021} using only  the original and ambiguous user queries $q_n$;
\item[SPLADE\_goldQuery] (kind of upper bound):  SPLADE using the manually rewritten query $q^*_n$;
\item[CQE] \cite{lin_contextualized_2021}, a state-of-the-art dense contextualized query representation learned using learning-to-rank on a dataset with pseudo-labels.
\end{inparadesc}
While the two former aim at evaluating how much our model captures contextual information, the latter is a TREC CAsT participant closely related to our work. 

To model answers when representing the query using $\qanswers$, we used two historical ranges (``\textbf{All}''  with $k=n-1$ answers and ``\textbf{Last}'' where we use only the last one, i.e. $k=1$) and three types of answer inputs:
\begin{inparadesc}
 \item[Answer] in which answers are the canonical answers;
 \item[Answer-Short] in which sentences are filtered as in the best performing TREC CAsT approach~\cite{LinTREC2020Notebook2021}. This allows for consistent input length, at the expense of losing information;
 \item[Answer-Long] As answers from CANARD are short (a few sentences extracted from Wikipedia -- contrarily to CAsT ones), we expand them to reduce the discrepancy between training and inference.
 For each sentence, we find the Wikipedia passage it appears in (if it exists in ORConvQA~\cite{qu_open-retrieval_2020}), and sample a short snippet of 3 adjacent sentences from it.  
\end{inparadesc}

Finally, we  also conducted ablation studies (on the best of the above variants) by modifying either the way to use the historical context or the training loss:
\begin{inparadesc}
\item[flatContext] a one-encoder version of our SPLADE approach in which we concatenate all information of the context to apply SPLADE to obtain a single representation of the query (instead of two representations $\qqueries_n$ and $\qanswers$ as in Equations 2 and 3) trained using a MSE loss function (Eq. \ref{eq:mse-loss}) since there are no more two representations.
\item[MSE] the version of our SPLADE approach trained with the MSE loss (Eq. \ref{eq:mse-loss}) instead of the proposed one (Eq. \ref{eq:full-loss});
\item[cosine] the version of our SPLADE approach trained with a cosine loss instead of the proposed loss (Eq. \ref{eq:full-loss}). The cosine loss is interesting because it is invariant to the scaling factor that preserves the document ordering (Eq.\ref{eq:rsv}).
\end{inparadesc}

\subsubsection{Second-stage ranking scenarios.} We consider different scenario for our  second-stage ranking model:
\begin{inparadesc}
    \item[T5Mono\_RawQuery] the T5Mono ranking model \cite{nogueira_document_2020} applied on raw queries;
    \item[T5Mono\_GoldQuery] the T5Mono ranking model applied on gold queries;
    \item[T5Mono\_CQR] the T5Mono ranking model applied on query reformulation generated with a pre-trained T5 (using the CANARD dataset);
    \item[\textbf{CoSPLADE\_[context]\_[number]}]: different versions of our second-stage ranking model input (Eq. \ref{eq:reranking:t5-contextual-prompt}), varying 1) the presence or absence of the past queries within the reformulation, and 2) the number $K$ of keywords identified as relevant by the first-stage ranker: 5, 10, 20.
\end{inparadesc}

\subsubsection{TREC participant baselines.} For each evaluation campaign (2020 and 2021), we also compare our model with the best, the median and the lowest TREC CAsT participants presented in the two overviews \cite{dalton_cast_2020,dalton_cast_2021}, where participants are ranked according to the nDCG@3 metric. Please note that we are not able to assess if results are significant since we report the effectiveness metrics presented in the TREC overview \cite{dalton_trec_2020,dalton_cast_2021}.


\section{Results}

\subsection{First-stage ranking effectiveness}

In this section, we focus on the first-stage ranking component of our CoSPLADE model. To do so, we experiment different scenarios aiming at evaluating the impact of the designed loss (Eq. \ref{eq:full-loss}) and the modeling/utility of evidence sources (Equations  \ref{eq:qrep-hist} and \ref{eq:qanswers}). Results of these different baselines and scenarios on the TREC CAsT 2021 dataset are provided in Table \ref{tab:firststage} -- similar trends are observed on CAsT 2020. We provide detailed results (at query level) in the GitHub repository. 

\begin{table}[t]
    \centering
    \begin{tabular}{|c|c|c|c|c|Hc|}
        \hline
         & Recall@500 & MAP@500 & MRR & nDCG@500 & nDCG@5 & nDCG@3 \\
        \hline
        \multicolumn{7}{|c|}{Baselines} \\
        \hline
        SPLADE\_rawQuery & 30.8$\pm$2.7 & 5.5$\pm$0.9 & 21.3$\pm$2.9 & 17.8$\pm$1.8 & 12.8$\pm$1.9 & 13.1$\pm$2.1 \\
        SPLADE\_goldQuery & 68.8$\pm$2.0 & 16.1$\pm$1.2 & 55.5$\pm$3.3 & 42.8$\pm$1.7 & 35.2$\pm$2.4 & 38.3$\pm$2.8\\
        CQE \cite{lin_-batch_2021} from \cite{dalton_cast_2021} & 79.1 & 28.9 & 60.3 & 55.7 & - &43.8 \\ 
        \hline
        \multicolumn{7}{|c|}{Effect of answer processing: CoSPLADE\_$\ldots$ } \\
        \hline
        AllAnswers & 79.5$\pm$2.2 & 28.8$\pm$1.7 & 61.7$\pm$3.1 & 55.3$\pm$2.0 & 44.1$\pm$2.6 & 46.5$\pm$2.9 \\
        AllAnswers-short & 72.8$\pm$2.6 & 25.7$\pm$1.9 & 54.4$\pm$3.3 & 49.5$\pm$2.3 & 38.6$\pm$2.7 & 40.1$\pm$3.0 \\
        AllAnswers-long & 80.4$\pm$2.1 & 29.3$\pm$1.8 & 62.0$\pm$3.2 & 55.6$\pm$2.1 & 46.3$\pm$2.7 & \textbf{48.9$\pm$3.0} \\
        LastAnswer & 83.4$\pm$2.0 & 31.2$\pm$1.8 & 61.8$\pm$3.1 & 58.1$\pm$2.0 & 46.0$\pm$2.7 & 47.4$\pm$3.0 \\
        LastAnswer-short & 79.2$\pm$2.2 & 28.1$\pm$1.8 & 61.4$\pm$3.3 & 54.3$\pm$2.1 & 44.7$\pm$2.7 & 46.4$\pm$3.0 \\
        \textbf{LastAnswer-long} &  \textbf{85.2$\pm$1.8} & \textbf{32.0$\pm$1.7} & \textbf{64.3$\pm$03.0} & \textbf{59.4$\pm$1.9} & 47.7$\pm$2.6 & 48.6$\pm$3.0 \\
        \hline
        \multicolumn{7}{|c|}{CoSPLADE\_LastAnswer-long variants} \\
        \hline 
        flatContext & 77.0$\pm$2.0 &26.0$\pm$2.0 & 55.0$\pm$3.0 & 52.0$\pm$2.0 & 41.0$\pm$3.0 & 42.0$\pm$3.0\\
        MSE loss & 70.9$\pm$2.4 & 21.6$\pm$1.7 & 48.7$\pm$3.4 & 45.2$\pm$2.3 & 34.9$\pm$2.8 & 36.9$\pm$3.1\\
        cosine loss & 70.4$\pm$2.5 & 22.6$\pm$1.7 & 52.5$\pm$3.3 & 46.9$\pm$2.2 & 37.5$\pm$2.7 & 39.0$\pm$3.0\\
        \hline
    \end{tabular}
    \caption{Effectiveness of different scenarios of our first-stage ranking model on the TREC CAsT 2021.}
    \label{tab:firststage}
    \vspace{-0.6cm}
\end{table}

In general, one can see that all variants of our approach (CoSPLADE\_* models) 
outperform the scenario applying the initial version of SPLADE on raw and, more importantly, gold queries.
This is very encouraging since this latter scenario might be considered an oracle, i.e. the query is manually disambiguated. Finally, we improve the results over CQE \cite{lin_contextualized_2021} for all the metrics -- showing that our simple learning mechanism, combined with SPLADE, allows for achieving SOTA performance. 

 \paragraph{Leveraging queries and answers history better contextualizes the current query.}
 The results of the flatContext scenario w.r.t. to the  SPLADE\_goldQuery  allow comparing the impact of evidence sources related to the conversation since they both use the same architecture (SPLADE). We can observe that it obtains better results than  SPLADE\_goldQuery  (e.g., 77 vs. 68.8 for the Recall@500 metric), highlighting the usefulness of context to better understand the information need.

\paragraph{More detailed answers perform better.}
Since answers are more verbose than questions, including them is more complex, and we need to study the different possibilities (CoSPLADE\_AllAnswers* and CoSPLADE\_LastAnswer*). One can see that: 1) trimming answers (*-short) into a few keywords is less effective than considering canonical answers, but 2) it might be somehow effective when combined with the associated Wikipedia passage (*-long). Moreover, it seems more effective to consider only the last answer rather than the whole set of answers in the conversation history\footnote{This might be due to the simple way to use past answers, i.e. Eq. \ref{eq:qanswers}, but all the other variations we tried did not perform better}. Taking all together, these observations highlight the importance of the way to incorporate information from answers into the reformulation process.

\paragraph{Dual query representation with asymmetric loss leverages sparse query representations.}
The results of the flatContext scenario show that considering at once past queries and answers perform better (compared to the MSE loss scenario which is directly comparable). However, if we separate the representations \emph{and} use an asymmetric loss function, the conclusion changes.
Moreover, the comparison of our best scenario CoSPLADE\_LastAnswer-long with a similar scenario trained by simply using a MSE or a cosine losses reveals the effectiveness of our asymmetric MSE (Equation \ref{eq:asymloss}). Remember that this asymmetric loss encourages the consideration of previous answers in the query encoding. This reinforces our intuition that the conversation context, and particularly verbose answers, is important for the conversational search task. It also reveals that the context should be included at different levels in  the architecture (input and loss). 

\subsection{Second-stage ranking effectiveness}

In this section, we rely on the CoSPLADE\_LastAnswer-long model as a first stage ranker, and evaluate different variants of the second-stage ranking method relying on the T5Mono model. For fair comparison, we also mention results obtained by a T5Mono ranking model applied on raw and gold queries, as well as query reformulated using a T5 generative model. Results on the TREC CAsT 2021 dataset are presented in Table \ref{tab:second-stage}.\\

\begin{table}[t]
    \centering
    \begin{adjustbox}{width=\textwidth}
    \begin{tabular}{|c|c|c|c|c|Hc|}
        \hline
         & Recall@500 & MAP@500 & MRR & nDCG@500 & nDCG@5 & nDCG@3 \\
        \hline
        \multicolumn{7}{|c|}{Baselines} \\
        \hline
        T5Mono\_RawQuery & 78.4$\pm$2.3 & 21.0$\pm$1.8 & 39.6$\pm$3.2 & 45.9$\pm$2.1 & 29$\pm$2.9 & 28.4$\pm$3.0\\
        T5Mono\_GoldQuery & 86.1$\pm$1.7 & 44.1$\pm$1.9 & 78.7$\pm$2.7 & 68.5$\pm$1.8 & 63.2$\pm$2.7 & 64.6$\pm$2.8\\
        T5Mono\_CQR &  80.4$\pm$2.2 & 30.0$\pm$1.9 & 58.2$\pm$3.4 & 55.3$\pm$2.1 & 43.6$\pm$3.0 & 44.6$\pm$3.2\\
        \hline \multicolumn{7}{|c|}{coSPLADE-based second stage variants } \\
        \hline
        CoSPLADE\_NoContext\_5  & 84.3$\pm$1.8 & 31.7$\pm$2.0 & 61.6$\pm$3.3 & 58.1$\pm$2.0 & 45.5$\pm$2.8 & 45.9$\pm$3.1\\
        CoSPLADE\_NoContext\_10 & 83.1$\pm$1.9 & 32.0$\pm$1.7 & 66.0$\pm$3.1 & 59.1$\pm$1.9 & 48.5$\pm$2.6 & 49.8$\pm$2.9\\
        CoSPLADE\_NoContext\_20 & 84.8$\pm$1.7 & 33.4$\pm$1.8 & 66.0$\pm$3.0 & 60.4$\pm$1.8 & 47.4$\pm$2.6 & 49.6$\pm$2.9\\
        CoSPLADE\_Context\_5 & \textbf{85.0$\pm$1.7} & 35.0$\pm$1.8 & 68.4$\pm$3.0 & 61.7$\pm$1.9 & 51.5$\pm$2.6 & 51.5$\pm$02.9\\
        CoSPLADE\_Context\_10 & 84.8$\pm$1.7 & \textbf{36.5$\pm$1.9} & 67.8$\pm$3.1 & \textbf{63.0$\pm$1.9} & 52.0$\pm$2.7 & 53.3$\pm$3.1\\
        CoSPLADE\_Context\_20 & 84.9$\pm$1.7 & 35.5$\pm$1.8 & \textbf{69.8$\pm$3.0} & 62.2$\pm$1.9 & 51.9$\pm$2.6 &\textbf{ 54.4$\pm$2.9}\\
        \hline
    \end{tabular}
    \end{adjustbox}
    \caption{Effectiveness of different scenarios of our second-stage ranking model on TREC CAsT 2021.}
    \label{tab:second-stage}
    \vspace{-0.6cm}
\end{table}

The analysis of the CoSPLADE model variants allows to highlight different observations regarding the usability of the context and the number of keywords  added to the query. 
First, adding the previous questions to the current query in the prompt (i.e., ``Context'') seems to improve the query understanding and, therefore, positively impacts the retrieval effectiveness. For instance, when 5 keywords are added, the context allows reaching 51.5\% for the nDCG@3 against 45.9\% without context. 
Second, the performances tend to increase with the number of additional keywords, particularly for scenarios without context, which is sensible. This trend is less noticeable for the scenarios with context since the best 
scores are alternatively obtained 
considering either 10 or 20 keywords.
Notice that adding 10 or 20 keywords is more effective than  only considering 5 keywords (e.g. 54.4\% vs. 51.5\% for the nDCG@3 metric).  Thus, It seems that  keywords help to reformulate the initial information need but they can lead to saturation when they are too numerous and combined with other information.

By comparing the best model scenarios with the more basic scenarios applying the  T5Mono second-stage ranker on raw and gold queries, we can observe that our method allows improving the retrieval effectiveness regarding initial queries but is not sufficient for reaching the performance of T5Mono\_GoldQuery. However, results obtained when applying T5Mono on queries reformulated by T5 highlight that the contextualization of an initial query is a difficult task. Indeed, the T5Mono\_CQR scenario is less effective than the T5Mono\_GoldQuery one,
depending on the metrics, the scores differ from 6 to 20 points.

Moreover, it is interesting to notice that  the SPLADE model applied on raw and gold queries (first-stage ranking in Table \ref{tab:firststage}) obtains lower results than the T5Mono model on the same data (second-stage ranking in Table \ref{tab:second-stage}). It can be explained by  the different purposes of those architectures:
SPLADE is a sparse model focusing on query/document expansion  while T5Mono is particularly devoted to increase precision. However, it is worth noting that combining SPLADE and T5Mono as first and second-stage rankers reaches the highest effectiveness results in our experimental evaluation. This demonstrates
the effectiveness of  the CoSPLADE approach to both contextualize queries and effectively rank documents.


\begin{table}[t]
    \centering
    \begin{adjustbox}{width=\textwidth}
    \begin{tabular}{|c|c|c|c|c|Hc|}
        \hline
        TREC CAsT 2020          & Recall@1000 & MAP@1000 & MRR & nDCG@1000 & nDCG@5 & nDCG@3 \\
        \hline
        TREC Participant (best) & 63.3 & 30.2 & 59.3 & 52.6 & - & 45.8\\
        TREC Participant (median) & 52.1 & 15.1 & 42.2 & 36.4& - & 30.4\\
        TREC Participant (low) & 27.9 & 1.0 & 5.9 & 11.1 & - & 2.2\\
        \hline
        CoSPLADE & 82.4$\pm$2.0 & 26.9$\pm$1.5 & 58.1$\pm$2.9 & 54.2$\pm$1.8 & 41.2$\pm$2.4 & 44.0$\pm$2.7 \\
        \hline \hline
        TREC CAsT 2021 & Recall@500 & MAP@500 & MRR & nDCG@500 & nDCG@5 & nDCG@3 \\\hline
        TREC Participants 1 (best) & 85.0 & 37.6 & 67.9 & 63.6 & - & 52.6 \\
        TREC Participants 2 (median) & 36.4 & 17.6 & 53.4 & 33.6 & - & 37.7 \\
        TREC Participants 3 (low) & 58.9 & 7.6 & 27.0 & 31.4 & - & 15.4 \\
        \hline
        CoSPLADE & 84.9$\pm$1.7 & 35.5$\pm$1.8 & 69.8$\pm$3 & 62.2$\pm$1.9 & 51.99$\pm$2.6 & 54.4$\pm$2.9\\
        \hline
    \end{tabular}
    \end{adjustbox}
    \caption{TREC CAsT 2020 and 2021 performances regarding participants}
    \label{table:previous-years-comparison}
    \vspace{-0.6cm}
\end{table}

\subsection{Effectiveness compared to TREC CAsT participants}

We finally compare our approach with TREC CAsT participants for the 2020 and 2021 evaluation campaigns (table \ref{table:previous-years-comparison}).
For both years, we reached very close or better performances than the best participants.
Indeed, CoSPLADE outperforms the best TREC participant for the 2020 evaluation campaign regarding Recall@1000 and nDCG@1000. For 2021, our model obtains better results than the best one for the MRR and nDCG@3 metrics. 
For both years, the best participant is the h2oloo team \cite{LinTREC2020Notebook2021,dalton_cast_2021} where they use query reformulation techniques, using T5. Our results suggest that our approach focusing on a sparse first-stage ranking model allows combining the benefit of query expansion and document ranking in a single model that eventually helps the final reranking step. In other words, simply rewriting the query without performing a joint learning document ranking can hinder the overall performance of the search task.

\subsection{Efficiency}

The first stage ranker is based on SPLADE, a sparse retrieval model and is therefore efficient. Detailed analysis in terms of FLOPs can be found in \cite{formal_splade_v2}. As our loss does not include any regularization loss to preserve sparsity, it is interesting to look at how they evolve compared to SPLADE.
%
The average number of non-zero entries increases only for our AllAnswers variants (from around 60 for gold/raw to 80), showing that in the future we might benefit from controlling sparsity to improve the CoSPLADE efficiency. Interestingly, the LastAnswers variants have a slightly higher sparsity and perform better.


\vspace{-0.2cm}
\section{Conclusion}
\vspace{-0.2cm}
In this paper, we have shown how a sparse retrieval neural IR model, namely SPLADE~\cite{formal_splade_v2}, could be leveraged together with a lightweight learning process to obtain a state-of-the-art first-stage ranker. We further showed that this first-stage ranker could be used to provide context to the second-stage ranker, leading to results comparable with the best-performing systems. Future work may explore strategies to better capture the information from the context or to explicitly treat user feedback present in the evaluation dataset. We also envision to evaluate our approach on other conversational QA datasets, such as  CoQA \cite{ReddyCM19}, OR-ConvQA \cite{Qu0CQCI20}, or ConvMix \cite{ChristmannRW22}.

\section{Acknowledgements}
We would like to thank ANR for supporting this work under the following grants: ANR JCJC SESAMS (ANR-18-CE23-0001) and ANR COST (ANR-18-CE23-0016-003).  This work was granted access to the HPC resources of IDRIS under the allocation AD011012847.

\bibliographystyle{splncs04}
\bibliography{main}

\begin{thebibliography}{10}
\providecommand{\url}[1]{\texttt{#1}}
\providecommand{\urlprefix}{URL }
\providecommand{\doi}[1]{https://doi.org/#1}

\bibitem{aliannejadi_harnessing_2020}
Aliannejadi, M., Chakraborty, M., Ríssola, E.A., Crestani, F.: Harnessing
  evolution of multi-turn conversations for effective answer retrieval pp.
  33--42. \doi{10.1145/3343413.3377968}, \url{http://arxiv.org/abs/1912.10554}

\bibitem{waterloo2020}
Arabzadeh, N., Clarke, C.L.A.: Waterlooclarke at the trec 2020 conversational
  assistant track (2020)

\bibitem{ChristmannRW22}
Christmann, P., Roy, R.S., Weikum, G.: Conversational question answering on
  heterogeneous sources. In: Amig{\'{o}}, E., Castells, P., Gonzalo, J.,
  Carterette, B., Culpepper, J.S., Kazai, G. (eds.) {SIGIR} '22: The 45th
  International {ACM} {SIGIR} Conference on Research and Development in
  Information Retrieval, Madrid, Spain, July 11 - 15, 2022. pp. 144--154. {ACM}
  (2022). \doi{10.1145/3477495.3531815},
  \url{https://doi.org/10.1145/3477495.3531815}

\bibitem{waterloo2019clarke}
Clarke, C.L.A.: Waterlooclarke at the {TREC} 2019 conversational assistant
  track. In: Voorhees, E.M., Ellis, A. (eds.) Proceedings of the Twenty-Eighth
  Text REtrieval Conference, {TREC} 2019, Gaithersburg, Maryland, USA, November
  13-15, 2019. {NIST} Special Publication, vol.~1250. National Institute of
  Standards and Technology {(NIST)} (2019),
  \url{https://trec.nist.gov/pubs/trec28/papers/WaterlooClarke.C.pdf}

\bibitem{Culpepper0S18}
Culpepper, J.S., Diaz, F., Smucker, M.D.: Research frontiers in information
  retrieval: Report from the third strategic workshop on information retrieval
  in lorne {(SWIRL} 2018). {SIGIR} Forum  \textbf{52}(1),  34--90 (2018).
  \doi{10.1145/3274784.3274788}, \url{https://doi.org/10.1145/3274784.3274788}

\bibitem{dalton_cast_2020}
Dalton, J., Xiong, C., Callan, J.: {CAsT} 2020: The conversational assistance
  track overview p.~10

\bibitem{dalton_trec_2019}
Dalton, J., Xiong, C., Callan, J.: {TREC} {CAsT} 2019: The conversational
  assistance track overview \url{http://arxiv.org/abs/2003.13624}

\bibitem{dalton_trec_2020}
Dalton, J., Xiong, C., Callan, J.: {TREC} {CAsT} 2019: The conversational
  assistance track overview \url{http://arxiv.org/abs/2003.13624}

\bibitem{dalton_cast_2021}
Dalton, J., Xiong, C., Callan, J.: {TREC} {CAsT} 2021: {The} {Conversational}
  {Assistance} {Track} {Overview} p.~7 (2021)

\bibitem{ElgoharyCanYouUnpack2019}
Elgohary, A., Peskov, D., Boyd-Graber, J.: Can {You} {Unpack} {That}?
  {Learning} to {Rewrite} {Questions}-in-{Context}. In: Proceedings of the 2019
  {Conference} on {Empirical} {Methods} in {Natural} {Language} {Processing}
  and the 9th {International} {Joint} {Conference} on {Natural} {Language}
  {Processing} ({EMNLP}-{IJCNLP}). pp. 5918--5924. Association for
  Computational Linguistics, Hong Kong, China (Nov 2019).
  \doi{10.18653/v1/D19-1605}, \url{https://aclanthology.org/D19-1605}

\bibitem{formal_splade_v2}
Formal, T., Lassance, C., Piwowarski, B., Clinchant, S.: From {Distillation} to
  {Hard} {Negative} {Sampling}: {Making} {Sparse} {Neural} {IR} {Models} {More}
  {Effective}. In: Proceedings of the 45th {International} {ACM} {SIGIR}
  {Conference} on {Research} and {Development} in {Information} {Retrieval}.
  pp. 2353--2359. {SIGIR} '22, Association for Computing Machinery, New York,
  NY, USA (Jul 2022). \doi{10.1145/3477495.3531857},
  \url{http://doi.org/10.1145/3477495.3531857}

\bibitem{formal_splade_2021}
Formal, T., Piwowarski, B., Clinchant, S.: {SPLADE}: {Sparse} {Lexical} and
  {Expansion} {Model} for {First} {Stage} {Ranking}. In: Proceedings of the
  44th {International} {ACM} {SIGIR} {Conference} on {Research} and
  {Development} in {Information} {Retrieval}. pp. 2288--2292. {SIGIR} '21,
  Association for Computing Machinery, New York, NY, USA (Jul 2021).
  \doi{10/gm2tf2}, \url{https://doi.org/10.1145/3404835.3463098}

\bibitem{Hofsttter2020ImprovingEN}
Hofst{\"a}tter, S., Althammer, S., Schr{\"o}der, M., Sertkan, M., Hanbury, A.:
  Improving efficient neural ranking models with cross-architecture knowledge
  distillation. ArXiv  \textbf{abs/2010.02666} (2020)

\bibitem{10.1145/3404835.3462891}
Hofst\"{a}tter, S., Lin, S.C., Yang, J.H., Lin, J., Hanbury, A.: Efficiently
  teaching an effective dense retriever with balanced topic aware sampling. In:
  Proceedings of the 44th International ACM SIGIR Conference on Research and
  Development in Information Retrieval. p. 113–122. SIGIR '21, Association
  for Computing Machinery, New York, NY, USA (2021).
  \doi{10.1145/3404835.3462891}, \url{https://doi.org/10.1145/3404835.3462891}

\bibitem{khattab_colbert_2020}
Khattab, O., Zaharia, M.: {ColBERT}: Efficient and effective passage search via
  contextualized late interaction over {BERT}
  \url{http://arxiv.org/abs/2004.12832}

\bibitem{KrasakisZeroshotQueryContextualization2022b}
Krasakis, A.M., Yates, A., Kanoulas, E.: Zero-shot {Query} {Contextualization}
  for {Conversational} {Search}. In: Proceedings of the 45th {International}
  {ACM} {SIGIR} {Conference} on {Research} and {Development} in {Information}
  {Retrieval}. pp. 1880--1884. {SIGIR} '22, Association for Computing
  Machinery, New York, NY, USA (Jul 2022). \doi{10.1145/3477495.3531769},
  \url{https://doi.org/10.1145/3477495.3531769}

\bibitem{kumar_making_nodate}
Kumar, V., Callan, J.: Making information seeking easier: An improved pipeline
  for conversational search p.~10

\bibitem{albert2020Lan}
Lan, Z., Chen, M., Goodman, S., Gimpel, K., Sharma, P., Soricut, R.: {ALBERT:}
  {A} lite {BERT} for self-supervised learning of language representations. In:
  8th International Conference on Learning Representations, {ICLR} 2020, Addis
  Ababa, Ethiopia, April 26-30, 2020. OpenReview.net (2020),
  \url{https://openreview.net/forum?id=H1eA7AEtvS}

\bibitem{lin_contextualized_2021}
Lin, S.C., Yang, J.H., Lin, J.: Contextualized query embeddings for
  conversational search \url{http://arxiv.org/abs/2104.08707}

\bibitem{lin_-batch_2021}
Lin, S.C., Yang, J.H., Lin, J.: In-batch negatives for knowledge distillation
  with tightly-coupled teachers for dense retrieval. In: Proceedings of the 6th
  Workshop on Representation Learning for {NLP} ({RepL}4NLP-2021). pp.
  163--173. Association for Computational Linguistics.
  \doi{10.18653/v1/2021.repl4nlp-1.17},
  \url{https://aclanthology.org/2021.repl4nlp-1.17}

\bibitem{LinTREC2020Notebook2021}
Lin, S.C., Yang, J.H., Lin, J.: {TREC} 2020 {Notebook}: {CAsT} {Track}. Tech.
  rep., TREC (Dec 2021)

\bibitem{lin_multi-stage_2021}
Lin, S.C., Yang, J.H., Nogueira, R., Tsai, M.F., Wang, C.J., Lin, J.:
  Multi-stage conversational passage retrieval: An approach to fusing term
  importance estimation and neural query rewriting
  \url{http://arxiv.org/abs/2005.02230}

\bibitem{query2020lin}
Lin, S., Yang, J., Nogueira, R., Tsai, M., Wang, C., Lin, J.: Query
  reformulation using query history for passage retrieval in conversational
  search. CoRR  \textbf{abs/2005.02230} (2020),
  \url{https://arxiv.org/abs/2005.02230}

\bibitem{mele_finding_2021}
Mele, I., Muntean, C.I., Nardini, F.M., Perego, R., Tonellotto, N.: Finding
  {Context} through {Utterance} {Dependencies} in {Search} {Conversations}.
  Tech. rep. (2021)

\bibitem{nogueira_document_2020}
Nogueira, R., Jiang, Z., Pradeep, R., Lin, J.: Document ranking with a
  pretrained sequence-to-sequence model. In: Findings of the Association for
  Computational Linguistics: {EMNLP} 2020. pp. 708--718. Association for
  Computational Linguistics. \doi{10.18653/v1/2020.findings-emnlp.63},
  \url{https://www.aclweb.org/anthology/2020.findings-emnlp.63}

\bibitem{qu_open-retrieval_2020}
Qu, C., Yang, L., Chen, C., Qiu, M., Croft, W.B., Iyyer, M.: Open-retrieval
  conversational question answering pp. 539--548.
  \doi{10.1145/3397271.3401110}, \url{http://arxiv.org/abs/2005.11364}

\bibitem{open2020qu}
Qu, C., Yang, L., Chen, C., Qiu, M., Croft, W.B., Iyyer, M.: Open-retrieval
  conversational question answering. In: Proceedings of the 43rd International
  ACM SIGIR Conference on Research and Development in Information Retrieval. p.
  539–548. SIGIR '20, Association for Computing Machinery, New York, NY, USA
  (2020). \doi{10.1145/3397271.3401110},
  \url{https://doi.org/10.1145/3397271.3401110}

\bibitem{Qu0CQCI20}
Qu, C., Yang, L., Chen, C., Qiu, M., Croft, W.B., Iyyer, M.: Open-retrieval
  conversational question answering. In: Huang, J.X., Chang, Y., Cheng, X.,
  Kamps, J., Murdock, V., Wen, J., Liu, Y. (eds.) Proceedings of the 43rd
  International {ACM} {SIGIR} conference on research and development in
  Information Retrieval, {SIGIR} 2020, Virtual Event, China, July 25-30, 2020.
  pp. 539--548. {ACM} (2020). \doi{10.1145/3397271.3401110},
  \url{https://doi.org/10.1145/3397271.3401110}

\bibitem{bert2019chen}
Qu, C., Yang, L., Qiu, M., Croft, W.B., Zhang, Y., Iyyer, M.: Bert with history
  answer embedding for conversational question answering. In: Proceedings of
  the 42nd International ACM SIGIR Conference on Research and Development in
  Information Retrieval. p. 1133–1136. SIGIR'19, Association for Computing
  Machinery, New York, NY, USA (2019). \doi{10.1145/3331184.3331341},
  \url{https://doi.org/10.1145/3331184.3331341}

\bibitem{qu2019attentive}
Qu, C., Yang, L., Qiu, M., Zhang, Y., Chen, C., Croft, W.B., Iyyer, M.:
  Attentive history selection for conversational question answering. In:
  Proceedings of the 28th ACM International Conference on Information and
  Knowledge Management. pp. 1391--1400 (2019)

\bibitem{reddy-etal-2019-coqa}
Reddy, S., Chen, D., Manning, C.D.: {C}o{QA}: A conversational question
  answering challenge. Transactions of the Association for Computational
  Linguistics  \textbf{7},  249--266 (2019). \doi{10.1162/tacl_a_00266},
  \url{https://aclanthology.org/Q19-1016}

\bibitem{ReddyCM19}
Reddy, S., Chen, D., Manning, C.D.: Coqa: {A} conversational question answering
  challenge. Trans. Assoc. Comput. Linguistics  \textbf{7},  249--266 (2019).
  \doi{10.1162/tacl\_a\_00266}, \url{https://doi.org/10.1162/tacl\_a\_00266}

\bibitem{santhanam-etal-2022-colbertv2}
Santhanam, K., Khattab, O., Saad-Falcon, J., Potts, C., Zaharia, M.:
  {C}ol{BERT}v2: Effective and efficient retrieval via lightweight late
  interaction. In: Proceedings of the 2022 Conference of the North American
  Chapter of the Association for Computational Linguistics: Human Language
  Technologies. pp. 3715--3734. Association for Computational Linguistics,
  Seattle, United States (Jul 2022). \doi{10.18653/v1/2022.naacl-main.272},
  \url{https://aclanthology.org/2022.naacl-main.272}

\bibitem{vakulenko_question_2021}
Vakulenko, S., Longpre, S., Tu, Z., Anantha, R.: Question rewriting for
  conversational question answering. In: Proceedings of the 14th {ACM}
  International Conference on Web Search and Data Mining. pp. 355--363. {ACM}.
  \doi{10.1145/3437963.3441748},
  \url{https://dl.acm.org/doi/10.1145/3437963.3441748}

\bibitem{voskarides_ilps_2019}
Voskarides, N., Li, D., Panteli, A., Ren, P.: {ILPS} at {TREC} 2019
  conversational assistant track p.~4

\bibitem{voskarides_query_2020}
Voskarides, N., Li, D., Ren, P., Kanoulas, E., de~Rijke, M.: Query resolution
  for conversational search with limited supervision pp. 921--930.
  \doi{10.1145/3397271.3401130}, \url{http://arxiv.org/abs/2005.11723}

\bibitem{waterloo2021}
Yan, X., Clarke, C.L.A., Arabzadeh, N.: Waterlooclarke at the trec 2021
  conversational assistant track (2021)

\bibitem{Yang2019QueryAA}
Yang, J.H., Lin, S.C., Wang, C.J., Lin, J.J., Tsai, M.F.: Query and answer
  expansion from conversation history. In: TREC (2019)

\bibitem{yu_few-shot_2020}
Yu, S., Liu, J., Yang, J., Xiong, C., Bennett, P., Gao, J., Liu, Z.: Few-shot
  generative conversational query rewriting
  \url{http://arxiv.org/abs/2006.05009}

\bibitem{ZamaniConversationalInformationSeeking2022}
Zamani, H., Trippas, J.R., Dalton, J., Radlinski, F.: Conversational
  {Information} {Seeking} (Jan 2022). \doi{10.48550/arXiv.2201.08808},
  \url{http://arxiv.org/abs/2201.08808}, arXiv:2201.08808 [cs]

\end{thebibliography}

\end{document}